\RequirePackage{ifpdf}
\ifpdf 
\documentclass[pdftex]{sigma}
\else
\documentclass{sigma}
\fi

\numberwithin{equation}{section}

\def\BI{{\rm 1\!l}}

\def\starwedge{\stackrel{\star}{\wedge}}

\begin{document}

\allowdisplaybreaks

\renewcommand{\thefootnote}{$\star$}

\renewcommand{\PaperNumber}{019}

\FirstPageHeading

\ShortArticleName{Emergent Abelian Gauge Fields from Noncommutative Gravity}

\ArticleName{Emergent Abelian Gauge Fields \\
 from Noncommutative Gravity\footnote{This paper is a
contribution to the Special Issue ``Noncommutative Spaces and Fields''. The
full collection is available at
\href{http://www.emis.de/journals/SIGMA/noncommutative.html}{http://www.emis.de/journals/SIGMA/noncommutative.html}}}

\Author{Allen STERN}

\AuthorNameForHeading{A. Stern}

\Address{Department of Physics and Astronomy, University of Alabama,
Tuscaloosa, Al 35487, USA}
\Email{\href{mailto:astern@bama.ua.edu}{astern@bama.ua.edu}}

\ArticleDates{Received December 30, 2009, in f\/inal form February 14, 2010;  Published online February 18, 2010}

\Abstract{We  construct exact solutions to noncommutative gravity following the formulation of Chamseddine and show that  they are in general accompanied by Abelian gauge f\/ields which are
 f\/irst order in the noncommutative scale. This provides a mechanism for  gene\-ra\-ting cosmological electromagnetic f\/ields in  an expanding space-time background, and also leads to multipole-like f\/ields surrounding black holes.  Exact  solutions to noncommutative Einstein--Maxwell theory can give rise   to  f\/irst order corrections
to the  metric tensor, as well as to the electromagnetic f\/ields.  This leads to f\/irst order shifts in the horizons of charged black holes.}


\Classification{83D05; 53D55; 81T75}

\renewcommand{\thefootnote}{\arabic{footnote}}
\setcounter{footnote}{0}

\section{Introduction}

A number of arguments point to a breakdown of the Riemannian description of space-time at the Planck scale.
In particular, it was been argued by Doplicher et al.~\cite{Doplicher:1994zv}
 that a new set of uncertainty relations for distance and time measurements are relevant at the Planck energy scale.
 These uncertainty relations are naturally realized within the context of noncommutative geometry,
 where space-time coordinates are replaced by noncommuting operators.
 This indicates that noncommutative space-time geometry may be the appropriate setting for Planck scale gravity.
 Among other things,  it has the potential of resolving the point singularities which plague general relativity,
 since the standard notion of `points' is absent in such a setting\footnote{This issue was addressed in the
 simpler context of a noncommutative deformation of electrodynamics
  where it was shown that noncommutative ef\/fects tend to screen the Coulomb singularity at leading order
  in the deformation~\cite{Stern:prl,Stern:2008wi}.}.

  Corrections to the metric tensor due to noncommutative geometry have been computed for a number of solutions to
  general relativity~\cite{Pinzul:2005ta,Chaichian:2007we,[us]}.
  However, so far, the prospects of observing these corrections are not  very promising.
  This is due to the smallness of the noncommutative scale\footnote{Conservative estimates from Hydrogen spectra put limits on the noncommutative scale at around a few GeV~\cite{Stern:prl,Stern:2008wi}.},
  plus the fact  that the previously obtained corrections to the metric tensor are quadratic in the noncommutative scale~\cite{Calmet:2006iz,Banerjee:2007th}.  Moreover, the procedure
used to obtain the above corrections has some ambiguity.  It does not involve solving the noncommutative analogue of  Einstein equations, which are not unique.  Rather, one maps  known solutions of general relativity to the noncommutative theory using the Seiberg--Witten map~\cite{Seiberg:1999vs}.  A noncommutative analogue of the metric tensor can be def\/ined  in order to then give a physical interpretation of the results, but the def\/inition is not unique.

  Here we shall examine exact solutions to
  noncommutative gravity, and show that they  can give rise to ef\/fects which are linear in the noncommutative scale when additional gauge f\/ields are present in the theory. Such additional f\/ields are in fact
  necessary if one follows the approach to noncommutative gravity of
  Chamseddine~\cite{Chamseddine:2000si,Chamseddine:2004si}.  The latter is based on the  gauge theory formulation of
  general relativity.  Because the relevant gauge group for gravity is non unitary,
  one needs to enlarge the gauge group in order to pass to the noncommutative theory. This requires adding at least two
  Abelian gauge f\/ields to the theory.  Their coupling to point particles was shown to be identical to that of electromagnetism (although one of them is associated with a noncompact direction of the gauge group)~\cite{Stern:2009jh}.
  We shall examine several standard solutions of general relativity, and show that the additional gauge f\/ields pick up f\/irst
  order contributions when going to the noncommutative gravity  theory.  Starting with the Robertson--Walker metric, this then provides a mechanism for  generating  cosmological electromagnetic f\/ields. Conversely, if  an electromagnetic f\/ield is present in  a solution to the Einstein--Maxwell equations, it can induce f\/irst order corrections in the metric
  tensor.

 Dif\/ferent formulations of noncommutative gravity have been proposed.  Among them is the approach of  Wess and  collaborators~\cite{Aschieri:2005yw},  where the full dif\/feomorphism group is preserved.  \mbox{Exact} solutions to the latter have been recently discussed in~\cite{Schupp:2009pt,Ohl:2009pv} which preserve the isometries of the corresponding commutative solution.  (See also~\cite{Dolan:2006hv}.)
    Actually, exact solutions can be obtained for a  wide class of
 noncommutative gravity theories.
  For this one only needs to require a)~that the commutative solutions are associated with some isometry group ${\cal I}$
  and b)~that the noncommutative  f\/ield equations   are deformations of the commutative equations
   obtained in a~canonical fashion.  Here the term `canonical'  means that  pointwise products appearing in the commutative theory
 get replaced by   star products.  As discussed below,
we shall utilize the Seiberg--Witten map~\cite{Seiberg:1999vs} in order to make contact
with a theory of space-time.
 (This dif\/fers from~\cite{Schupp:2009pt,Ohl:2009pv}.)
  Because the Seiberg--Witten map  can be problematic for
some noncommutative manifolds, we shall specialize to embeddings of the Moyal--Weyl plane, with its realization in terms of the
  Groenewold--Moyal  star product~\cite{groe,moy}.   It is def\/ined in terms of dimensionfull space-time independent  parameters $\Theta^{\mu\nu}$, setting the noncommutative scale.
  One can  choose $\Theta^{\mu\nu}$ such that the  Groenewold--Moyal  star  product acts trivially (i.e., as a pointwise product) between
  any two functions associated with the isometry group ${\cal I}$.  Noncommutative f\/ield equations evaluated for such functions are then
  equivalent to commutative f\/ield equations, and so solutions to commutative gravity with  isometry group ${\cal I}$
 must  also be solutions to the noncommutative theory.

    The f\/inal task is then to give a space-time interpretation to the noncommutative solutions.   Unlike in previous treatments, we do not  try to def\/ine  a noncommutative analogue  of the metric tensor for this purpose.   We instead  use the Seiberg--Witten map~\cite{Seiberg:1999vs} to map
   noncommutative solutions  onto the space of commutative f\/ield conf\/igurations, where the standard metric tensor, as well as other familiar quantities of
  general relativity, can be employed. The Seiberg--Witten map can then potentially generate f\/irst order shifts in both the metric tensor and the Abelian gauge f\/ields of the commutative solution.  Because  $\Theta^{\mu\nu}$ are space-time independent, these shifts will in general break various space-time symmetries associated with the commutative solution.   We  carry out the  procedure for three examples: the f\/lat expanding universe,  the Schwarzschild black hole and the Reissner--Nordstr\"om black hole.  First order Abelian gauge f\/ields emerge in the  f\/irst  two examples, and f\/irst order corrections to the metric tensor result in the last example.

 The organization of this article is as follows.  We review the noncommutative $GL(2,C)$ gauge theory  of~\cite{Chamseddine:2004si} in Section~\ref{section2}, along with its Seiberg--Witten map from the commutative theory in Section~\ref{section3}. A prescription for f\/inding exact solutions to the noncommutative Einstein equations is described in Section~\ref{section4}.  We apply the procedure to the case of the f\/lat expanding universe in Section~\ref{section5} and the Schwarzschild metric in Section~\ref{section6}.
  We generalize the procedure in Section~\ref{section7} in order to f\/ind  exact solutions to the noncommutative Einstein--Maxwell system  and apply it to the case of Reissner--Nordstr\"om black hole in Section~\ref{section8}.  Concluding remarks are made in Section~\ref{section9}. Expressions for the  noncommutative  gauge variations, the curvature and torsion and the Seiberg--Witten map are given for  component one forms in the appendix.

\section[Noncommutative $GL(2,C)$ gauge theory of gravity]{Noncommutative $\boldsymbol{GL(2,C)}$ gauge theory of gravity}\label{section2}

The standard gauge theory formalism for gravity  \cite{Utiyama:1956sy,Kibble:1961ba,Hehl:1976kj,Chamseddine:2005td}
  is expressed in terms of vierbeins~$e^a_\mu$ and spin connections $\omega^{ab}_\mu=-\omega^{ba}_\mu$.    $a,b,\ldots=0,1,2,3$ denote Lorentz indices which are raised and lowered with the f\/lat metric tensor $\eta_{ab}={\rm diag}(-1,1,1,1)$, while $\mu,\nu,\ldots$ are space-time indices which are lowered with the space-time metric
\begin{gather}
 g_{\mu\nu}= e^a_{\mu}e^b_{\nu}\eta_{ab}  .\label{mtrcnendom}
 \end{gather}
For convenience we use spinor notation of  \cite{Isham:1972br} and introduce the  matrix-valued one forms
\begin{gather*}
 e = e^a_\mu \gamma_a dx^\mu,\qquad   \omega =\tfrac 12 \omega^{ab}_\mu\sigma_{ab} dx^\mu  ,
 \end{gather*}
with $\{\gamma_a,\gamma_b\}=2\eta_{ab}\BI$ ($\BI$ denotes the unit matrix) and $SL(2,C)$ generators $\sigma_{ab}= -\frac i4 [\gamma_a,\gamma_b]$.  $ [\;,\;]$ and $\{\;,\;\}$ denote the matrix commutator and anticommutator, respectively.
Inf\/initesimal variations $\delta_\lambda$ of   $\omega$  and $e$ induced by  local $SL(2,C)$ transformations
are given  by
\begin{gather}
 \delta_\lambda \omega = d\lambda +i[\omega,\lambda], \qquad  \delta_\lambda e = i[e,\lambda] , \label{lcllrnz}
 \end{gather} where $  \lambda =\frac 12  \lambda^{ab}\sigma_{ab}$ is inf\/initesimal.  Such transformations leave the metric tensor (\ref{mtrcnendom}) invariant.
The spin curvature $R$ and torsion $T$ are constructed from  $\omega$ and $e$  according to
 \begin{gather*}
   R= \tfrac 12 R^{ab}_{\mu\nu} \sigma_{ab}dx^\mu\wedge dx^\nu= d\omega + i\omega \wedge \omega,  \\
  T = T^a_{\mu\nu}\gamma_adx^\mu\wedge dx^\nu= de +i(\omega\wedge e + e\wedge \omega ) ,
 \end{gather*}
 the latter of which is set to zero in the usual metric description of general relativity.
 $R$ and $T$  satisfy the Bianchi identities
\begin{gather*}
 dR = i(R\wedge \omega -\omega\wedge R),\\
   dT= i(R\wedge e-e\wedge R+T\wedge \omega-\omega \wedge T).
   \end{gather*}

 The canonical procedure to go from a commutative to a noncommutative theory is to
replace the pointwise products of the commutative theory by  star products, more specif\/ically  Groenewold--Moyal star products~\cite{groe,moy}, given by
\begin{gather}
 \star = \exp \biggl\{ \frac {i}2 \Theta^{\mu\nu}\overleftarrow{
  \partial_\mu}\;\overrightarrow{ \partial_\nu} \biggr\}  ,\label{gmstr}
  \end{gather}
    where
 $\Theta^{\mu\nu}=-\Theta^{\nu\mu}$ are constant matrix elements denoting the noncommutativity parameters and
  $\overleftarrow{
  \partial_\mu}$ and $\overrightarrow{ \partial_\mu}$ are left and right
derivatives,
respectively, with respect to  some space-time coordinates $x^\mu$. If we call $\hat A,\hat B,\ldots$ the noncommutative analogues of the matrix-valued forms $A,B,\ldots$ of the commutative theory, then the  matrix-valued star commutator
$[\hat A,\hat B]_\star = $  $\hat A\star\hat B - \hat B\star\hat A$ replaces the  matrix commutator $[A,B]$  when we go to the noncommutative theory.  Gauge theories based on non unitary groups, such as $SL(2,C)$, do not have a straightforward  generalization to noncommutative gauge theories.  (See \cite{Jurco:2000ja,Bonora:2000td}.)  For this note that the $SL(2,C)$ algebra is not realized with the star commutator.  This follows because the star commutator of matrix valued forms can be decomposed in terms of commutators, as well as anticommutators, of the algebra generators.  The star commutators of forms proportional to $\sigma_{ab}$ and $\gamma_a$ then generate the entire Clif\/ford algebra. Such star commutators will appear upon trying to lift the  gauge variations in (\ref{lcllrnz}) to noncommutative variations.  The commutators  $ [\omega,\lambda]$ and $[e,\lambda]$ in (\ref{lcllrnz}) get replaced by
\begin{gather}
 [\hat\omega,\hat\lambda]_\star =\tfrac 18 \{\hat \omega^{ab}, \hat \lambda^{cd}\}_\star \, [\sigma_{ab},\sigma_{cd}] + \tfrac 18 [\hat \omega^{ab}, \hat \lambda^{cd}]_\star\, \{\sigma_{ab},\sigma_{cd}\}, \label{clsron}\\
 [\hat e,\hat\lambda]_\star = \tfrac 14 \{\hat e^{a}, \hat \lambda^{bc}\}_\star \, [\gamma_{a},\sigma_{bc}] + \tfrac 14 [\hat e^{a}, \hat \lambda^{bc}]_\star\,  \{\gamma_{a},\sigma_{bc}\}  ,\label{clsrtw}
 \end{gather}
where $  \hat \omega =\frac 12 \hat\omega^{ab}_\mu\sigma_{ab} dx^\mu$, $ \hat e = \hat e^a_\mu \gamma_a dx^\mu$ and $\hat \lambda =\frac 12 \hat \lambda^{ab}\sigma_{ab}$ are  the noncommutative analogues of the  connection one form  $\omega$, the vierbein one form $e$ and the gauge parameters $\lambda$, respectively, and $\{\;,\;\}_\star$ denotes the star anticommutator,
$\{\hat \alpha,\hat \beta\}_\star =\hat \alpha\star\hat \beta + \hat \beta\star\hat \alpha.$  Using the Groenewold--Moyal star, $\{\hat \alpha,\hat \beta\}_\star$ is real for any two real-valued functions $\hat \alpha$ and $\hat \beta$,   while  $[\hat \alpha,\hat \beta]_\star$ is imaginary.  The anticommutator  $\{\sigma_{ab},\sigma_{cd}\}$ appearing in (\ref{clsron}) is a linear combination of $\gamma_5$ and the unit matrix~$\BI$.  Closure therefore requires that we  enlarge the gauge group to $GL(2,C)$ in order to include these two generators.  Following \cite{Chamseddine:2004si} one  then introduces $GL(2,C)$ connections $\hat {\cal A}$ and  gauge parameters $\hat \Lambda$ which can be decomposed as
\begin{gather*}
 \hat{\cal A} =\hat \omega+ \hat A^{(1)}\BI+ i\hat A^{(2)}\gamma_5,\qquad  \hat \Lambda =\hat \lambda+\hat \alpha^{(1)}\BI + i\hat \alpha^{(2)} \gamma_5 ,
 \end{gather*}  where $\hat A^{(S)}$, $S=1,2$,  are Abelian connection one forms,
$ \hat  A^{(S)} =\hat  A^{(S)}_\mu dx^\mu $, and $\hat \alpha^{(S)}$  are functions on space-time.  The anticommutator $ \{\sigma_{ab},\gamma_c\}  $ appearing in~(\ref{clsrtw}) is a linear combination of $\gamma_5\gamma_c$, and  so closure  requires that one generalizes  the vierbein one forms $\hat e$ to
\begin{gather*}
  \hat{\cal E}= \hat e +\hat f ,\qquad  \hat f=\hat f^a_\mu \gamma_5\gamma_a dx^\mu .
  \end{gather*}
A consistent set of noncommutative $GL(2,C)$ gauge variations can then be def\/ined according~to
\begin{gather}
 \delta_{\hat \Lambda}\hat {\cal  A }= d\hat \Lambda +i[\hat{\cal  A},\hat \Lambda]_\star, \qquad   \delta_{\hat \Lambda}\hat  {\cal E} = i[\hat {\cal E},\hat\Lambda]_\star.\label{ncvlcllrnz}
 \end{gather}
These variations  are decomposed in terms of component f\/ields $\hat \omega^{ab}$, $\hat A^{(S)}$, $ \hat e^a$ and $ \hat f^a $ in (\ref{cmpsncgtrns}) in the appendix.
The noncommutative $GL(2,C)$ curvature $\hat{\cal F}$ and torsion $\hat {\cal T}$ are given  by
\begin{gather*}
  \hat {\cal F}= d\hat{\cal  A} + i\hat {\cal  A}\starwedge  \hat {\cal A} = \tfrac 12 \hat R^{ab}\sigma_{ab} + \hat F^{(1)} \BI +i\hat F^{(2)} \gamma_5, \\
 \hat {\cal T}= d\hat {\cal E} +i(\hat {\cal A}\starwedge\hat {\cal E} + \hat {\cal E}\starwedge \hat {\cal A} )  = \hat T^a\gamma_a +\hat U^a\gamma_5\gamma_a ,
 \end{gather*}
where $\starwedge$ denotes an exterior product   where
the usual pointwise product  between components of the
forms replaced by the Groenewold--Moyal star product.
The components~$\hat R^{ab}$,~$ \hat F^{(S)}$,~$ \hat T^a$ and~$\hat U^a$ of the  noncommutative curvature and torsion two forms are given in (\ref{crvtrncmpnts}) in the appendix.
 The Bianchi identities now read
\begin{gather*}
 d\hat {\cal F} =i(\hat {\cal F}\starwedge \hat{\cal  A} -\hat {\cal A}\starwedge\hat {\cal F}),\\
  d\hat {\cal T}=i(\hat {\cal F}\starwedge \hat {\cal E}-\hat {\cal E}\starwedge \hat {\cal F}+\hat {\cal T}\starwedge \hat {\cal A}-\hat {\cal A }\starwedge \hat {\cal T}).
 \end{gather*}

\section[First order Seiberg-Witten map]{First order Seiberg--Witten map}\label{section3}

The  Seiberg--Witten map~\cite{Seiberg:1999vs}  can be used to  map  the noncommutative gravity theory back to a commutative gravity theory, where quantities like the metric tensor can be straightforwardly def\/ined. However, the commutative gravity theory here is not   the standard $SL(2,C)$ gauge theory description of gravity, but rather it is a $GL(2,C)$ gauge theory.  The latter is def\/ined in terms of the  one forms ${\cal A}={\cal A}_\mu dx^\mu=\omega+  A^{(1)}\BI+ i A^{(2)}\gamma_5$ and ${\cal E}={\cal E}_\mu dx^\mu=  e^a \gamma_a + f^a \gamma_5\gamma_a $, which are the commutative analogues of $\hat {\cal A}$ and $ \hat {\cal E}$, respectively.  The commutative $GL(2,C)$ gauge variations   are given by
\begin{gather*}
 \delta_\Lambda {\cal A} = d \Lambda +i[{\cal  A}, \Lambda], \qquad   \delta_\Lambda {\cal E} = i[ {\cal E},\Lambda], 
\end{gather*}
 with inf\/initesimal parameters $\Lambda =\frac12  \lambda^{ab}\sigma_{ab}+ \alpha^{(1)}\BI + i \alpha^{(2)} \gamma_5$.
In terms of component f\/ields $\omega^{ab}$, $ A^{(S)}$, $  e^a$ and $  f^a $, this  means
\begin{gather}
 \delta_\Lambda \omega^{ab} = d \lambda^{ab} +  \omega^{ac}\lambda_c^{\;\;b} -  \omega^{bc}\lambda_c^{\;\;a} , \nonumber\\
\delta_\Lambda A^{(S)}  =  d \alpha^{(S)}  ,\qquad S=1,2,\nonumber\\
  \delta_\Lambda e^a =  e^b \lambda_b^{\;\;a}         +2 f^a \alpha^{(2)},\nonumber\\
 \delta_\Lambda f^a  =   f^b \lambda_b^{\;\;a}          +2 e^a \alpha^{(2)}.
\label{cmpsgtrns}
\end{gather}
Equation (\ref{mtrcnendom}) is not suitable as a metric tensor for this theory since it is not invariant under the full set of $GL(2,C)$ transformations.
The following  $GL(2,C)$ invariant metric tensor was  given in~\cite{Stern:2009jh}:
\begin{gather}
 {\tt g}_{\mu\nu} = \tfrac 14 \, {\rm tr}\, {\cal E}_\mu {\cal E}_\nu =e^a_{\mu}e_{a\nu} -f^a_{\mu}f_{a\nu}.\label{gltcmtrc}
\end{gather}
 Because this gravity theory possesses two vierbein f\/ields $e^a$
and $f^a$, one can construct two sets of torsion two-forms
\begin{gather}
  T^a=d e^a +\tfrac 12[ \omega^{ab},  e_b],
\qquad   U^a=d f^a +\tfrac 12[ \omega^{ab},  f_b] ,\label{cmtvtandU}
\end{gather}
which are the commutative analogues of $\hat T^a$ and  $\hat U^a$.

The Seiberg--Witten map relates the noncommutative degrees of freedom in $\hat {\cal A}_{\mu}$ and  $ \hat {\cal E} _\mu$, along with transformation parameters $\hat \Lambda$, to their commutative counterparts $ {\cal A}_{\mu}$,  $  {\cal E} _\mu$ and $ \Lambda$, and is  def\/ined such that  gauge transformations in the commutative theory induce  gauge transformations  in the corresponding noncommutative theory. Up to f\/irst order in $\Theta^{\mu\nu}$ it can be given~by
\begin{gather}
   \hat {\cal A}_{\mu}={\cal A}_{\mu}+\tfrac 14 \Theta^{\rho\sigma} \{ {\cal A}_\rho ,\partial_\sigma {\cal A}_\mu+ {\cal F}_{\sigma\mu}\}+{\cal O}\big(\Theta ^2\big),\nonumber\\
   \hat {\cal E} _\mu
 ={\cal E} _\mu+\tfrac 12\Theta^{\rho\sigma} \big\{  {\cal  A}_\rho,\partial_\sigma{\cal E}_\mu +\tfrac i2 [{\cal A}_\sigma,{\cal E}_\mu] \big\}+{\cal O}\big(\Theta ^2\big),\nonumber\\
 \hat \Lambda
 =\Lambda +\tfrac 14\Theta^{\rho\sigma} \{   {\cal A}_\rho,\partial_\sigma \Lambda \}+{\cal O}\big(\Theta ^2\big).
 \label{frstrdrSW}
 \end{gather}
The resulting f\/irst order expressions   for the component one forms $\hat \omega^{ab}$, $\hat A^{(S)}$, $\hat e^a$ and $\hat f^a$ are given in (\ref{swutfo}) of the appendix\footnote{For a related Seiberg--Witten map, see \cite{Marculescu:2008gw}.}.

\section{Exact solutions to noncommutative gravity}\label{section4}

Here we do not specify any particular dynamics for  noncommutative gravity, but only demand that the f\/ield equations are deformations of the standard Einstein equations, with deformation parameters $\Theta^{\mu\nu}$, and are obtained in the canonical way, i.e., by replacing pointwise products with Groenewold--Moyal star products. Such a procedure is, of course, not unique.
One can nevertheless write down exact solutions for the noncommutative theory, as we describe below.

Say a solution to  (commutative) gravity is associated with some nontrivial isometry group~${\cal I}$,
which is generated by $n$ Killing vectors $K_\alpha$, $\alpha=1,\dots,n$.
Denote by $V_A$, $A=1,\dots,4-n$, the remaining independent vectors  normal to~$K_\alpha$.   Now def\/ine $\Theta^{\mu\nu} $ such that
\begin{gather}
 \Theta^{AB} V_A \otimes V_B =0.
\label{tvavb}
\end{gather}
 If all $K_\alpha$ vanishes on  a set of functions ${\cal F}$, then the  bivectors $\Theta^{\alpha B} K_\alpha \otimes V_B$ and $\Theta^{\alpha \beta} K_\alpha \otimes K_\beta$  vanish on ${\cal F}\otimes {\cal F}$.
Moreover, from~(\ref{tvavb}) it follows that
 $\Theta^{\mu\nu}\partial_\mu \otimes \partial_\nu $ vanishes
on ${\cal F}\otimes {\cal F}$.  From the def\/inition of the
Groenewold--Moyal star~(\ref{gmstr}), one gets that the star product
of any pair of functions in ${\cal F}$ is identical to its corresponding pointwise product. So when evaluated on elements of ${\cal F}$, any noncommutative f\/ield equation, obtained using the canonical procedure, reduces to the corresponding commutative f\/ield equation.
Therefore solutions to commutative gravity with isometry group~${\cal I}$
 must  also be solutions to the noncommutative theory.
The former solutions are  specif\/ied  by $SL(2,C)$ vierbein and spin connection one forms, which we denote respectively by $e^a_{(0)}=e^a_{(0)\mu}dx^\mu$ and $\omega^a_{(0)}=\omega^a_{(0)\mu}dx^\mu$.  Then the corresponding noncommutative solution is given by
\begin{gather}
  \hat e^a_\mu= e^a_{(0)\mu}, \qquad \hat \omega^{ab}_\mu=\omega^{ab}_{(0)\mu}, \qquad  \hat A^{(S)}_\mu=\hat f^a_\mu=0. \label{ncslntzordr}
  \end{gather}
 The above procedure can be adapted for obtaining exact solutions can be applied to a wide class of noncommutative  theories, i.e., those obtained in the canonical way.  Exact solutions were found in \cite{Schupp:2009pt,Ohl:2009pv} to the noncommutative gravity theory of~\cite{Aschieri:2005yw}.  They were also used  to f\/ind exact solutions to noncommutative electrodynamics~\cite{Stern:2008wi}.

In order to give a physical interpretation to the exact noncommutative solution (\ref{ncslntzordr}), one can use (\ref{frstrdrSW}) to map the solution to the commutative theory.   The  $SL(2,C)$ vierbeins $  e^a_{(0)\mu}$ and spin connections $ \omega^{ab}_{(0)\mu}$ pick up no   f\/irst order corrections  from the Seiberg--Witten map.  On the other hand, the map does generate nonvanishing f\/irst order results for $ A^{(S)}_\mu$ and $  f^a_\mu$:
\begin{gather}
  A^{(1)}_\mu =  -\tfrac1{16} \Theta^{\rho\sigma} [ \omega_{(0)ab}]_{\rho} \bigl(\partial_\sigma \omega^{ab}_{(0)\mu} +R^{ab}_{(0)\sigma\mu}\bigr)+{\cal O}\big(\Theta ^2\big),\nonumber\\
   A^{(2)}_\mu = \tfrac1{32} \Theta^{\rho\sigma}  \epsilon_{abcd}\omega^{ab}_{(0)\rho} \bigl(\partial_\sigma \omega^{cd}_{(0)\mu} +R^{cd}_{(0)\sigma\mu}\bigr) +{\cal O}\big(\Theta ^2\big),\nonumber\\
   f^a_\mu = \tfrac 14 \Theta^{\rho\sigma}\epsilon^a_{\;\;bcd}\omega^{bc}_{(0)\rho }  \bigl(\partial_\sigma e^d_{(0)\mu} +\frac 12 \omega^{dg}_{(0)\sigma } e_{(0)g\mu} \bigr)+{\cal O}\big(\Theta ^2\big),
\label{gnrtablngfs}
\end{gather} where $R_{(0)}$ is the Lorentz curvature for the solution.
Since the $GL(2,C)$ invariant metric tensor~(\ref{gltcmtrc}) is quadratic in  $f^a_\mu$, it receives no f\/irst order corrections.  This is consistent with the usual result that noncommutative gravity corrections to the space-time metric tensor are second order in the noncommutativity parameter~\cite{Calmet:2006iz,Banerjee:2007th}.  This, however, would  {\it not} be  the case for solutions of the Einstein--Maxwell equations which are associated with a nonzero Abelian curvature, since then $ \hat A^{(S)}$ is nonzero.  Therefore an Abelian gauge f\/ield can generate f\/irst order corrections to the metric tensor.  We show this in Section~\ref{section7} and give the example of the Reissner--Nordstr\"om  black hole solution in Section~\ref{section8}.  The induced   $ f^a_\mu$ in~(\ref{gnrtablngfs}) can lead to a f\/irst order  $GL(2,C)$ torsion  $U^a=d f^a +{\cal O}(\Theta^2)$, but its physical meaning  is not immediately  evident.

In the next two sections we carry out the above procedure to f\/ind exact solutions of the noncommutative theory for the case of the f\/lat expanding universe and the Schwarzschild black hole and  we obtain the leading order induced Abelian gauge f\/ields in these two examples.  For the former example, $any$ choice for the constants $\Theta^{\mu\nu}$ satisf\/ies (\ref{tvavb}) when expressed in terms of comoving coordinates, while for  static solutions, (\ref{tvavb}) is always satisf\/ied for the case of space-time noncommutativity.  In addition to~(\ref{tvavb}), there is the requirement  that the
generators of the  star product algebra are  self-adjoint operators.  This is the case for all the examples which follow.

 \section{Flat expanding universe}\label{section5}

 The  invariant measure for a f\/lat expanding universe is
  \begin{gather}
   ds^2_{(0)} = -dt^2 + a(t)^2 dx_i dx_i  ,\qquad  i=1,2,3 ,\label{feuimsr}
   \end{gather}
 where $x_i$ span $R^3$ and  def\/ine comoving coordinates. Equation~(\ref{feuimsr}) gives a solution to the Einstein equations with stress-energy tensor of the form
 \begin{gather}
 {\tt T}^{SE}_{\;\mu\nu}= {\rm diag}\left(  \rho(t) , a(t)^2p(t), a(t)^2p(t), a(t)^2p(t)\right),\label{sefrfeu}
 \end{gather}
  where $\rho(t)$ and $p(t)$ are the proper energy density and pressure, respectively, and are related to~$a(t)$. Vierbein and spin connection one forms consistent with the  zero torsion condition of general relativity can be given by
\begin{gather*}
 e^0_{(0)} =  dt,\qquad e^i_{(0)}  = a(t)dx_i, \\
 \omega^{0i}_{(0)} = {\dot a(t)}dx_i, \qquad  \omega^{ij}_{(0)}  =  0.
\end{gather*}

Here all spatial directions   correspond to Killing vectors $K_i=\frac \partial{\partial x^i}$. Since only one independent vector $V_0 =\frac\partial{\partial t}$ remains, the condition (\ref{tvavb}) is satisf\/ied for any choice of $\Theta^{\mu\nu}$. Therefore, (\ref{ncslntzordr})  is an exact solution to the noncommutative gravity equations (obtained using the canonical procedure) with stress-energy tensor (\ref{sefrfeu}), for any choice of $\Theta^{\mu\nu}$.  We can thus choose space-space noncommutativity
\begin{gather}
 [x_i,x_j]_\star =\Theta^{ij}  ,\label{ssnctvt}
 \end{gather}
or  time-space noncommutativity
\begin{gather}
[t,x_i]_\star  = \Theta^{0i}, \label{stnctvt}
\end{gather} or both.
 (\ref{ssnctvt}) and (\ref{stnctvt}) break  the three dimensional rotation symmetry.  This symmetry breaking  will not appear at f\/irst order in the metric tensor, but does appear in the electric f\/ield at f\/irst order.  As was remarked previously~\cite{Ohl:2009pv}, these  commutation relations  apply for the comoving  coordinates, while the commutators between `physical' spatial coordinates $y_i$  involve the scale factor,~$y_i=a(t)x_i$.  As a result,  the `physical' noncommutative scale will become extremely small at earlier scales, and one would expect that this prevents the noncommutative ef\/fects  from becoming too large at  earlier times.  This could be desirable, since such ef\/fects  break rotational invariance. Despite this, as we argue below, the f\/ields might become signif\/icant during the inf\/lation era.

 Next  we substitute into (\ref{gnrtablngfs}).
 A nonvanishing f\/irst order potential $ A^{(1)}_\mu$   only arises for the case of time-space noncommutativity~(\ref{stnctvt})
 \begin{gather*}   A^{(1)}_0 = {\cal O}\big(\Theta ^2\big),\qquad   A^{(1)}_i = -\tfrac 14  \Theta^{0i}  \dot a(t) \ddot a(t) +{\cal O}\big(\Theta ^2\big) ,
  \end{gather*}
  leading to a uniform   electric f\/ield at f\/irst order,  directed along $\Theta^{0i}$,
 \begin{gather}
    E_i^{(1)} =  -\tfrac 14\Theta^{0i} \big(\ddot a(t)^2+\dot a(t) a^{(3)}(t)\big)+{\cal O}\big(\Theta ^2\big). \label{Eqstfl}
 \end{gather}
 A nonvanishing f\/irst order  potential $ A^{(2)}_\mu$  only arises for the case of space-space noncommutativity~(\ref{ssnctvt})
 \begin{gather*}
    A^{(2)}_0 = {\cal O}\big(\Theta ^2\big),\qquad A^{(2)}_i = \tfrac 18 \epsilon_{ijk}\Theta^{jk}  \dot a(t)^3 +{\cal O}\big(\Theta ^2\big) ,
\end{gather*}
leading to a uniform  electric f\/ield at f\/irst order, in the $\epsilon_{ijk}\Theta^{jk}$ direction,
 \begin{gather}
    E_i^{(2)} = \tfrac 38 \epsilon_{ijk}\Theta^{jk}\dot a(t)^2 \ddot a(t)+{\cal O}\big(\Theta ^2\big).
    \label{Essf2}
\end{gather}
Since the associated magnetic f\/ields vanish, it follows that the standard Maxwell equations in vacuum are not satisf\/ied.  Rather, the dynamics for the  Abelian gauge f\/ields is governed by a~deformed Maxwell action, and as a result, magnetic f\/ields are shielded by the noncommutative vacuum.

If one applies the results (\ref{Eqstfl}) and (\ref{Essf2}) to the universe at the current time one gets incredibly tiny electric f\/ields, since  they go like powers of the Hubble  parameter (not to mention the small noncommutative scale~$\Theta$).  The f\/ields grow when we evolve   back in time.  Yet, they remain insignif\/icant even if we go all the way back to the beginning of the radiation dominated era.
For the radiation dominated  era we can   apply
\begin{gather}
 a(t) = a_{rm}\left(\frac t{t_{rm}}\right)^{1/2},\label{radsclfctr}
   \end{gather}
   where $a_{rm}\sim 3\times 10^{-4}$ and $t_{rm}\sim 5\times 10^4 \;$yr, respectively, are the scale factor and time of the radiation-matter equality.  Substituting into  (\ref{Eqstfl}) and (\ref{Essf2}) gives
\begin{gather*}
   E_i^{(1)} =-\frac1{16}{\Theta^{0i}}  \frac{a_{rm}^2}{t_{rm }t^3}+{\cal O}\big(\Theta ^2\big), \qquad
  E_i^{(2)} = -\frac 3{128} \epsilon_{ijk}\Theta^{jk}   \frac{a_{rm}^3}{t_{rm} ^{3/2}t^{5/2}}+{\cal O}\big(\Theta ^2\big).
\end{gather*}
If  we take known upper limits~\cite{Stern:prl,Stern:2008wi} for the time-space and space-space noncommutativity scale  of $\Theta \lesssim {\rm GeV}^{-2}$,  and set $t=t_{rm}$, we get
$ |E^{(1)}| \lesssim   (10^{-67} \; {\rm eV})^2$ and  $|E^{(2)}| \lesssim
(10^{-69} \, {\rm eV})^2$.   These scales are nowhere near those claimed needed for the primordial (magnetic) f\/ield ($10^{-15}$--$10^{-25}\; {\rm eV}^2$) to  seed an amplif\/ication process which can then produce the currently observed  intergalactic magnetic f\/ields \cite{Davis:1999bt}\footnote{The application of noncommutative physics to the primordial magnetic f\/ield problem was suggested earlier in~\cite{Mazumdar:2000jc}, but the f\/ield there originates from the matter content of the universe, and  is not generated from the Robertson--Walker background, as is the case here.}.
If we assume that (\ref{radsclfctr}) is valid all the way back to, say  $t\sim 10^{-30}$~sec (and we take the previous limit for $\Theta$),  we get  $ |E^{(1)}| \lesssim (10^{-3} \; {\rm eV})^2$ and  $|E^{(2)}| \lesssim
(10^{-16}\;  {\rm eV})^2$. These energy scales are still far below the radiation energy ($\sim 10^6\; {\rm GeV}$) at the time $t\sim 10^{-30}\;{\rm  sec}$, and so the f\/ields are not expected to play a role.

 On the other hand, the f\/ields $ |E^{(1)}|$ and $ |E^{(2)}|$ can be very large during a prior exponential inf\/lationary era.   (Note that the f\/ields depend on derivatives of $a(t)$, which are not continuous across the transition from the inf\/lationary era to the radiation era if one assumes that the exponential inf\/lation was suddenly switched of\/f.  The f\/ields therefore undergo a discontinuous jump across the transition.)   Upon taking $a(t)\propto e^{Ht}$,
both  $ |E^{(1)}| $ and     $ |E^{(2)}| $ go like $H^4$ times the noncommutative scale.  Since here $H$ is associated with a large energy scale, these f\/ields could  be signif\/icant during inf\/lation.

\section{Schwarzschild solution}\label{section6}

The Schwarzschild   invariant measure may be written as
\begin{gather*}
 ds^2_{(0)}=\bigl(-1+\alpha(r)^2\bigr) dt^2 + \left(  \frac{\alpha(r)^2}{1-\alpha(r)^2}\hat x_i\hat x_j +\delta_{ij}\right) dx_i dx_j
 ,\qquad  \alpha(r)^2= \frac{r_s}r  ,
 \end{gather*}
where the spatial coordinates  $x_i$ again span $R^3$. $r=\sqrt{x_ix_i}$ is the radial coordinate, $\hat x_i=x_i/r$ are unit vectors and $r_s=2GM$ is the Schwarzschild radius.  A consistent set of vierbein one forms $e_{(0)}$  is
\begin{gather}
 e^0_{(0)} = d t + \frac{\alpha(r) dr}{1-\alpha(r)^2},
\qquad
 e^i_{(0)} =\alpha(r)\hat x_i e^0_{(0)}    + d x_i. \label{ezroschwz}
 \end{gather}
   The torsion vanishes with the following assignment for the spin connections
   \begin{gather}
    \omega^{i0}_{(0)} = d\bigl( \alpha(r) \hat x_i\bigr) + \alpha(r)\alpha'(r) \hat x_i  e^0_{(0)},\qquad \omega^{ij}_{(0)} =0 , \label{omegazroschwz}
   \end{gather}
   where the prime denotes dif\/ferentiation in $r$.  Equations (\ref{ezroschwz}) and (\ref{omegazroschwz}) are valid both inside and outside the event horizon and lead to zero torsion and zero curvature (for $r>0$).

   As with all static solutions, the Schwarzschild solution  has a time-like Killing vector $K_0 =\frac\partial{\partial t}$. Now we identify the  spatially directed vectors  with $V_i=\frac \partial{\partial x^i}$. From~(\ref{tvavb}), we  get an exact solution of the noncommutative gravity equations when all space-space components of  $\Theta^{\mu\nu}$ vanish.   So here we only  consider time-space noncommutativity~(\ref{stnctvt}).
  The Groenewold--Moyal star product  when acting between $t$-independent functions reduces to the pointwise product, and so if the noncommutative gravity equations  are obtained in the canonical way,  they are equivalent to the commutative gravity equations when evaluated for static f\/ields. Equation~(\ref{ncslntzordr}) is then an exact solution of the noncommutative f\/ield equations with time-space noncommutativity.

   Now  substitute into~(\ref{gnrtablngfs})  to obtain the induced commutative  Abelian gauge potentials $ A^{(1)}_\mu$ and $ A^{(2)}_\mu$ at leading order.
 Here we f\/ind that the  former leads to an electrostatic f\/ield and the latter leads to a magnetostatic f\/ield originating from the black hole.  The vector potential $ A^{(1)}_i$ is   a pure gauge at f\/irst order and hence there is no associated  f\/irst order magnetic f\/ield  $B^{(1)}_i$, while
 \begin{gather*}
  A^{(1)}_0 =-\frac {\Theta^{0i}r_s^2 \hat x_i}{8r^5} +{\cal O}\big(\Theta ^2\big)  ,
  \end{gather*}
   resulting in the electrostatic f\/ield
 \begin{gather}
  E^{(1)}_{i} = \frac {\Theta^{0j}r_s^2}{8}  \frac{\delta_{ij} -6\hat x_i\hat x_j}{r^6}+{\cal O}\big(\Theta ^2\big). \label{eonest}
  \end{gather}
 On the other hand, $A^{(2)}_0={\cal O}(\Theta ^2)$  and hence there is no associated f\/irst order  electric  f\/ield  $E^{(2)}_{i}$, while
 \begin{gather*}
  A^{(2)}_i=  \frac {\Theta^{0k}r_s^2 }{8r^5} \epsilon_{ijk}\hat x_j  +{\cal O}\big(\Theta ^2\big),
 \end{gather*} resulting in the magnetostatic f\/ield
 \begin{gather}
  B^{(2)}_{i} =  \frac {\Theta^{0j}r_s^2}{4}  \frac{2\delta_{ij} -3\hat x_i\hat x_j}{r^6} +{\cal O}\big(\Theta ^2\big).\label{btwost}
\end{gather}

 The results (\ref{eonest}) and (\ref{btwost}) break the rotational symmetry of the Schwarzschild solution and  resemble higher multipole f\/ields, with ${\Theta^{0i}r_s^2}$ playing a role analogous to that of a multipole moment.   As the  f\/ields fall of\/f as~$1/r^6$, they are very weak far from the black hole.   At the Schwarzschild radius   $r=r_s$, they both go like  $\sim \Theta^{0i}/r_s^4 $. Since  ${\Theta^{0i}}$ is constant in this theory, the moments of multiple black holes are aligned, which may lead to  a peculiar signature for microscopic black holes.

\section[Exact solutions to  noncommutative Einstein-Maxwell theory]{Exact solutions to  noncommutative Einstein--Maxwell theory}\label{section7}

It is straightforward to generalize the procedure of Section~\ref{section4} to f\/ind exact  solutions to the combined noncommutative Einstein--Maxwell system.  As before, we do not specify any particular dynamics for the  noncommutative theory, but only demand that the f\/ield equations are deformations of the standard Einstein--Maxwell equations,  obtained in the canonical way.  In the  examples in Sections~\ref{section5} and~\ref{section6},  solutions to the noncommutative Einstein equations generated  Abelian gauge f\/ields which were f\/irst order in~$\Theta$. Conversely, here we show that solutions to noncommutative Maxwell  equations can generate f\/irst order corrections to the metric tensor.  For this
we   shall require, as before, that~(\ref{tvavb}) is satisf\/ied, and that the generators of the star product algebra are self-adjoint operators.

Solutions to the  commutative Einstein--Maxwell equations are specif\/ied  by  $SL(2,C)$ vierbein and spin connection one forms, $ e_{(0)}^a$ and $\omega_{(0)}^{ab}$, along with an Abelian potential one form  $A_{(0)}=A_{(0)\mu}dx^\mu$.  Since $GL(2,C)$ gauge theory contains two Abelian gauge f\/ields, we can construct two dif\/ferent exact solutions to noncommutative Einstein--Maxwell theory.  Below we f\/irst assume that $\hat A^{(1)}$ satisf\/ies the noncommutative Maxwell equation and hence identify $A_{(0)}$ with   $\hat A^{(1)}$,  and then we consider the case  for $\hat A^{(2)}$.

1. $\hat A^{(1)}$ satisf\/ies the noncommutative Maxwell equation.
If (\ref{tvavb}) holds we have the exact noncommutative solution
\begin{gather}
  \hat e^a_\mu= e^a_{(0)\mu}, \qquad\hat \omega^{ab}_\mu=\omega^{ab}_{(0)\mu}, \qquad \hat A^{(1)}_\mu=A_{(0)\mu},\qquad A^{(2)}_\mu=\hat f^a_\mu=0. \label{aonencslntzordr}
\end{gather}
Upon applying the Seiberg--Witten map back to the commutative theory, one now can pick up  f\/irst order corrections to all of the $GL(2,C)$  f\/ields.  For  the two  vierbein f\/ields $e^a_\mu $ and $f^a_\mu $, one has
\begin{gather*}
   e^a_\mu = e^a_{(0)\mu} -\Theta^{\rho\sigma} A_{(0)\rho} \bigl(\partial_\sigma e^a_{(0)\mu} +\tfrac 12 \omega^{ag}_{(0)\sigma} e_{(0)g\mu} \bigr) +{\cal O}\big(\Theta ^2\big),\nonumber\\
  f^a_\mu =  \tfrac 14 \Theta^{\rho\sigma}\epsilon^a_{\;\;bcd}\omega^{bc}_{(0)\rho}  \bigl(\partial_\sigma e^d_{(0)\mu} +\tfrac 12 \omega^{dg}_{(0)\sigma} e_{(0)g\mu} \bigr)+{\cal O}\big(\Theta ^2\big).
  \end{gather*}
    The former can then give rise to f\/irst order corrections to the classical metric tensor $g_{(0)\mu\nu}$. After substituting the expression for $ e^a_\mu $ into (\ref{gltcmtrc}), one gets the simple result that
\begin{gather}
  {\tt g}_{\mu\nu}(x) =g_{(0)\mu\nu}(x)-\Theta^{\rho\sigma} A_{(0)\rho}(x) \partial_\sigma g_{(0)\mu\nu}+{\cal O}\big(\Theta ^2\big)
  =g_{(0)\mu\nu}\left(x+\Theta A_{(0)}\right) +{\cal O}\big(\Theta ^2\big). \label{frrdrcrmtr}
  \end{gather}
Since ${\tt g}_{\mu\nu}$ transforms under dif\/feomorphisms as a rank 2 tensor, the f\/irst order correction cannot be  removed by the  coordinate redef\/inition $x\rightarrow x -\Theta A_{(0)} $.
From~(\ref{tvavb}), the bivector  $\Theta^{\mu\nu}\partial_\mu\otimes\partial_\nu $ gets contributions from  $\Theta^{\alpha B} K_\alpha \otimes V_B$ and $\Theta^{\alpha \beta} K_\alpha \otimes K_\beta$.   However, only the former produces f\/irst order corrections to  $g_{(0)\mu\nu}$, and from~(\ref{frrdrcrmtr}), they are of the form $ \delta g_{(0)\mu\nu}=-\Theta^{\alpha B} A_{(0)\alpha} V_B g_{(0)\mu\nu}$.
 Such correction  generally break the space-time symmetry associated with the  metric tensor~$g_{(0)\mu\nu}$.

As before, the Seiberg--Witten map produces  f\/irst order terms in the Abelian gauge f\/ields.  Here one gets some additional contributions to $A^{(1)}_\mu$
\begin{gather}
A^{(1)}_\mu = A_{(0)\mu} -\tfrac12 \Theta^{\rho\sigma}\bigl( A_{(0)\rho} \bigl(\partial_\sigma A_{(0)\mu} +F_{(0)\sigma\mu}\bigr)\nonumber\\
\phantom{A^{(1)}_\mu =}{} +\tfrac18[ \omega_{(0)ab}]_{\rho} \bigl(\partial_\sigma \omega^{ab}_{(0)\mu} +R^{ab}_{(0)\sigma\mu}\bigr) \bigl)+{\cal O}\big(\Theta ^2\big)
 ,\label{ncmeai2}
\end{gather}
  where $F_{(0)}$ is the Abelian curvature for the solution.  $A^{(2)}_\mu$ has the same form as in (\ref{gnrtablngfs}).
There are now also f\/irst order correction to the spin connections
\begin{gather}
  \omega^{ab}_\mu = \omega^{ab}_{(0)\mu} -\tfrac 12\Theta^{\rho\sigma}\bigl(\omega^{ab}_{(0)\rho}\bigl(\partial_\sigma A_{(0)\mu} +F_{(0)\sigma\mu}\bigr) + A_{(0)\rho }\bigl(\partial_\sigma \omega^{ab}_{(0)\mu} +R^{ab}_{(0)\sigma\mu}\bigr) \bigr)+{\cal O}\big(\Theta ^2\big).
  \label{nccrtspncn}
\end{gather}
   This will result in nonvanishing torsion two forms  (\ref{cmtvtandU}) at f\/irst order.

2. $\hat A^{(2)}$ satisf\/ies the noncommutative Maxwell equation.   If (\ref{tvavb}) holds we  have the exact noncommutative solution
\begin{gather}
  \hat e^a_\mu= e^a_{(0)\mu}, \qquad\hat \omega^a_\mu=\omega^a_{(0)\mu}, \qquad \hat A^{(2)}_\mu=A_{(0)\mu},\qquad A^{(1)}_\mu=\hat f^a_\mu=0. \label{atwoencslntzordr}
  \end{gather}
Upon applying the Seiberg--Witten map back to the commutative theory, one gets
\begin{gather*}
 e^a_\mu = e^a_{(0)\mu} -\tfrac 14\Theta^{\rho\sigma}\epsilon^a_{\;\;bcd}\omega^{bc}_{(0)\rho } A_{(0)\sigma} e^d_{(0)\mu}+{\cal O}\big(\Theta ^2\big),\nonumber\\
    f^a_\mu = \tfrac 14 \Theta^{\rho\sigma}\epsilon^a_{\;\;bcd}\omega^{bc}_{(0)\rho}  \bigl(\partial_\sigma e^d_{(0)\mu} +\tfrac 12 \omega^{dg}_{(0)\sigma} e_{(0)g\mu} \bigr) +{\cal O}\big(\Theta ^2\big),
\end{gather*}
 which now produces no f\/irst order correction to $g_{(0)\mu\nu}$.  The  corrections to the Abelian gauge potentials and spin connections  in this case are given by
 \begin{gather*}
   A^{(1)}_\mu = -\tfrac12 \Theta^{\rho\sigma}\bigl(- A_{(0)\rho} \bigl(\partial_\sigma A_{(0)\mu} +F_{(0)\sigma\mu}\bigr)+\tfrac18[ \omega_{(0)ab}]_{\rho} \bigl(\partial_\sigma \omega^{ab}_{(0)\mu} +R^{ab}_{(0)\sigma\mu}\bigr)\bigl)+{\cal O}\big(\Theta ^2\big),\nonumber\\
  A^{(2)}_\mu = A_{(0)\mu} +\tfrac1{32} \Theta^{\rho\sigma} \epsilon_{abcd}\omega^{ab}_{(0)\rho}\bigl(\partial_\sigma \omega^{cd}_{(0)\mu} +R^{cd}_{(0)\sigma\mu}\bigr) +{\cal O}\big(\Theta ^2\big), \nonumber\\
    \omega^{ab}_\mu = \omega^{ab}_{(0)\mu} -\tfrac 14\epsilon^{ab}_{\;\;\;\;cd}\Theta^{\rho\sigma}\bigl( \omega^{cd}_{(0)\rho}\bigl(\partial_\sigma A_{(0)\mu} +F_{(0)\sigma\mu}\bigr)+ A_{(0)\rho}\bigl(\partial_\sigma \omega^{cd}_{(0)\mu} +R^{cd}_{(0)\sigma\mu}\bigr)\bigr)
 +{\cal O}\big(\Theta ^2\big).
\end{gather*}

Of course, a f\/inal possibility is that  the noncommutative Maxwell equation  involves both  $\hat A^{(1)}$ and  $\hat A^{(2)}$, i.e., a linear combination of $\hat A^{(1)}$ and  $\hat A^{(2)}$ satisf\/ies the noncommutative Maxwell equations.  The f\/irst order corrections to the commutative solution will then be a linear combination of those for the two cases, and  would in general lead to  f\/irst  order corrections to~$g_{(0)\mu\nu}$.

\section[Reissner-Nordstr\"om solution]{Reissner--Nordstr\"om solution}\label{section8}

We now use the results of the   previous section to obtain f\/irst order corrections  to the Reissner--Nordstr\"om metric tensor.  The  invariant  interval for the  Reissner--Nordstr\"om solution is
\begin{gather}
 ds^2_{(0)}= - \Delta(r) dt^2 + \Delta(r)^{-1} dr^2 + r^2 \big(d\theta^2 + \sin^2\theta d\phi^2\big) ,\label{rnnvrntmtr}
\\
 \Delta(r)=1 - \frac{2GM}r +\frac{GQ^2}{r^2}.\label{dlta}
 \end{gather}
Along with  the Coulomb  gauge f\/ield
\begin{gather*}
 F_{(0)} = \frac Q{r^2} dr\wedge dt ,
 \end{gather*}
(\ref{rnnvrntmtr}) and (\ref{dlta})  describe a black hole with charge $Q$ and mass $M$.
The metric tensor gives rise to two horizons at $r=r_\pm$,
\begin{gather}
 r_\pm=G \left(M\pm\sqrt{ M^2-\frac{ Q^2}G}\right). \label{clhrzRN}
 \end{gather}
Introducing coordinates $x_i$, $i=1,2,3$, spanning $R^3$, we can re-write the invariant measure to have metric components
\begin{gather*}
 g_{(0)00}=- \Delta(r) ,   \qquad g_{(0)ij}= \left(\Delta(r)^{-1}-1\right)\hat x_i \hat x_j +\delta_{ij},\qquad g_{(0)0i}=  0.
 \end{gather*}

As this is a static solution, the condition (\ref{tvavb}) is satisf\/ied for the case of time-space noncommutativity (\ref{stnctvt}).   Following the discussion in  Section~\ref{section7}, we can construct two exact solutions, (\ref{aonencslntzordr}) and (\ref{atwoencslntzordr}), of the noncommutative Einstein--Maxwell equations.  We take the former, since we want to obtain f\/irst order corrections to the metric tensor.
  For simplicity, choose only $\Theta^{03}$ nonzero and take $A_{(0)} = -\frac Qr dt$. Upon substituting into (\ref{frrdrcrmtr}) we obtain the new invariant measure
$ ds^2=ds^2_{(0)} +ds^2_{(1)}$, where the f\/irst order correction is
\begin{gather*}
 ds^2_{(1)} = -
\frac {\Theta^{03}Q}r\bigl( \Delta'(r)\cos\theta \bigl(dt^2 +  \Delta(r)^{-2} dr^2 \bigr) + 2\bigl(  \Delta(r)^{-1}-1 \bigr)\sin\theta \, drd\theta  \bigr)    ,
\end{gather*} the
prime denoting a derivative in $r$.  Thus
\begin{gather*}
 ds^2 =- \Sigma(r, \theta) dt^2 + \Sigma(r, \theta)^{-1} dr^2 + r^2\big(d \theta^2 + \sin^2\theta d\phi^2\big)\\
\phantom{ds^2 =}{}  -
\frac {2\Theta^{03}Q}r \bigl(  \Delta(r)^{-1}-1 \bigr)\sin\theta \;drd\theta +{\cal O}\big(\Theta ^2\big)  ,
\end{gather*} where
\begin{gather*}
 \Sigma(r, \theta) =
 \Delta(r)+ \frac {\Theta^{03}Q}r \Delta'(r)\cos \theta.
 \end{gather*}
 Rotation invariance is therefore broken at f\/irst order in $\Theta^{03}$.  The two horizons~(\ref{clhrzRN}) are  shifted  to $r=r_\pm+ \delta r_\pm$, where $\delta r_\pm$ is $\Theta^{03}-$dependent
 \begin{gather*}
  \delta r_\pm= -\frac{\Theta^{03}Q}{r_\pm}  \cos \theta +{\cal O}\big(\Theta ^2\big).
  \end{gather*}
 The f\/irst order corrections to the metric tensor vanish in the $Q\rightarrow 0$ limit, consistent with the  results obtained in Section~\ref{section6} for the Schwarzschild black hole.
 Additional roots of $\Sigma(r, \theta)$ may also occur, and they appear to be nontrivial. Noncommutative corrections to black hole horizons and their ef\/fect on the Hawking temperature have been previously computed in \cite{Nasseri:2005ji,Nozari:2006rt,Banerjee:2008gc,Banerjee:2008du}.  Here such ef\/fects will be f\/irst order~$\Theta$.

  Using (\ref{ncmeai2}), f\/irst order corrections will also result in the Abelian gauge  f\/ields\footnote{They will include contributions found in~\cite{Stern:2008wi}  in the absence of gravity.}, as well as in the spin connections (\ref{nccrtspncn}), and they   generalize the results of Section~\ref{section6}.

\section{Concluding remarks}\label{section9}

We have constructed exact solutions to noncommutative gravity and shown that  they can generate  Abelian gauge f\/ields which are
 f\/irst order in the noncommutative scale.  Conversely,  Abelian gauge f\/ields, if present in the solution, give rise to  f\/irst order corrections
to the  metric tensor.  In both cases, the ef\/fects are expected to break the space-time symmetries  associated with the solution.

For the case of black holes, we found that   both electrostatic and magnetostatic type  f\/ields
are  generated when ${\Theta^{0i}}\ne0$.  They fall of\/f as $1/r^6$ and
resemble higher multipole f\/ields, with ${\Theta^{0i}r_s^2}$  playing a role analogous
to the multipole moment. Since  ${\Theta^{0i}}$ is constant in this theory, the moments of multiple black holes are aligned,  which may lead to  a peculiar signature for the detection of microscopic black holes.

We got extremely  tiny electric-type f\/ields at the current time for the case of the f\/lat expanding universe. The f\/ields grow when we evolve   back in time.  If we go back to the  time of the radiation-matter equality, their strength is nowhere near that claimed needed for the primordial  f\/ield  to  seed an amplif\/ication process which can produce the currently observed  intergalactic magnetic f\/ields.  The energy scales of the f\/ields remain well below the radiation energy even if we go back to the beginning of the radiation era.  On the other hand,  the electric f\/ields could be signif\/icant  during an exponential inf\/lationary era.  We plan on addressing their ef\/fect on inf\/lationary models in the future.  One possibility is that they reach the critical strength for the creation of charged particle pairs from the vacuum.

The system of \cite{Chamseddine:2004si} which was studied here contains two Abelian gauge f\/ields.  We have entertained the notion that one of them may correspond to ordinary electromagnetism. Although from (\ref{cmpsgtrns}), the inf\/initesimal gauge variations of $A^{(1)}$ and  $A^{(2)}$ are identical in the commutative theory, $A^{(1)}$ is the natural candidate for the electromagnetic potential since it is associated with the compact direction in $GL(2,C)$. Here note that
 f\/inite (commutative) $GL(2,C)$ gauge transformations are of the form
\begin{gather*}
 {\cal A}\rightarrow \Omega {\cal A} \Omega^{-1} +i \Omega d\Omega^{-1}, \qquad  {\cal E}\rightarrow \Omega {\cal E} \Omega^{-1} ,\qquad  \Omega=e^{i\Lambda}, 
 \end{gather*}
where   $\Lambda =\frac12  \lambda^{ab}\sigma_{ab}+ \alpha^{(1)}\BI + i \alpha^{(2)} \gamma_5$  now denote f\/inite   parameters. Concerning
$A^{(2)}$, it will couple to the axial vector current when including fermions.  The possibility of generalizing the internal Abelian gauge symmetries of this theory to standard model symmetries is of further interest.

Other avenues of research  involve examining alternatives to (\ref{ssnctvt}) and~(\ref{stnctvt}), which are  associated with constant commutation relations  for the comoving  coordinates.  As was remarked in Section~\ref{section5} and in~\cite{Ohl:2009pv}, one would get larger noncommutative ef\/fects at earlier times if one can instead implement constant commutation relations  for the  `physical', rather than the comoving, spatial coordinates,  as this would introduce extra factors of $a(t)^{-1}$.  However, even after including such  factors,  the f\/ields may still be too small  to play a role as a primordial f\/ield which seeds an amplif\/ication process that can produce the currently observed  intergalactic magnetic f\/ields.

The electric f\/ields which emerge from this gravity theory play a passive role and only appear as a result of the Seiberg--Witten map to the commutative theory.  It is of interest to promote them to dynamical f\/ields.  They are then expected to contribute symmetry breaking terms to the stress-energy tensor ((\ref{sefrfeu}), for the example of the f\/lat expanding universe) and generate a~back reaction on the space-time geometry.  Abelian gauge f\/ields which are linear in~$\Theta$ will lead to second order  back reaction terms in the metric tensor.

\appendix

\section{Appendix}\label{appendixA}

Here we give expressions for the  noncommutative  gauge variations, the curvature and torsion and the Seiberg--Witten map in terms of the component one forms $\hat \omega^{ab}$, $\hat A^{(S)}$, $\hat e^a $ and $\hat f^a $.  From~(\ref{ncvlcllrnz}),
the  noncommutative $GL(2,C)$ gauge variations  can be written as
\begin{gather}
\delta_{\hat \Lambda}\hat \omega^{ab} = d\hat \lambda^{ab} +\tfrac12 \bigl(\{\hat \omega^{ac},\hat\lambda_c^{\;\;b}\}_\star - \{\hat \omega^{bc},\hat\lambda_c^{\;\;a}\}_\star \bigr) +i[\hat \omega^{ab},\hat\alpha^{(1)}]_\star +i[\hat A^{(1)},\hat \lambda^{ab}]_\star \nonumber\\
\phantom{\delta_{\hat \Lambda}\hat \omega^{ab} =}{}
+\tfrac i2\epsilon^{abcd}([\hat\omega_{cd},\hat \alpha^{(2)}]_\star+[\hat  A^{(2)},\hat\lambda_{cd}]_\star),\nonumber\\
 \delta_{\hat \Lambda}\hat A^{(1)} = d\hat \alpha^{(1)} +i[\hat A^{(1)},\hat \alpha^{(1)}]_\star-i[\hat  A^{(2)},\hat\alpha^{(2)}]_\star+\tfrac i8[\hat \omega^{ab},\hat\lambda_{ab}]_\star, \nonumber\\
\delta_{\hat \Lambda}\hat  A^{(2)} = d\hat \alpha^{(2)} +i[\hat A^{(1)},\hat \alpha^{(2)}]_\star-i[\hat  A^{(2)},\hat\alpha^{(1)}]_\star-\tfrac i{16}\epsilon^{abcd}[\hat \omega_{ab},\hat\lambda_{cd}]_\star,\nonumber\\
 \delta_{\hat \Lambda}\hat e^a =\tfrac 12\{ \hat e^b,\hat \lambda_b^{\;\;a}\}_\star -\tfrac i4 \epsilon^{abcd}[\hat f_b,\hat\lambda_{cd}]_\star +i[\hat e^a,\hat \alpha^{(1)}]_\star            +\{\hat f^a,\hat \alpha^{(2)}\}_\star, \nonumber
\\
  \delta_{\hat \Lambda}\hat f^a =\tfrac 12\{ \hat f^b,\hat \lambda_b^{\;\;a}\}_\star -\tfrac i4 \epsilon^{abcd}[\hat e_b,\hat\lambda_{cd}]_\star +i[\hat f^a,\hat \alpha^{(1)}]_\star            +\{\hat e^a,\hat \alpha^{(2)}\}_\star.\label{cmpsncgtrns}
\end{gather}
The components $\hat R^{ab}$, $ \hat F^{(S)}$, $ \hat T^a$ and $\hat U^a$ of the  noncommutative curvature and torsion two forms are given by
\begin{gather}
 \hat R^{ab}=d\hat\omega^{ab} +\tfrac 12 [\hat \omega^{ac},\hat\omega_c^{\;\;b}]_\star +i\{\hat\omega^{ab},\hat A^{(1)}\}_\star +\tfrac i2\epsilon^{abcd}\{\hat\omega_{cd},\hat A^{(2)}\}_\star , \nonumber\\
  \hat F^{(1)}=d\hat A^{(1)} + \tfrac i8 \hat\omega^{ab}\starwedge \hat\omega_{ab}+ i\hat A^{(1)}\starwedge\hat A^{(1)} -i\hat A^{(2)}\starwedge\hat A^{(2)},\nonumber\\
  \hat F^{(2)}=d\hat A^{(2)}-\tfrac i {16} \epsilon_{abcd}\hat\omega^{ab}\starwedge \hat\omega^{cd}
+i\{\hat A^{(1)},\hat A^{(2)}\}_\star, \nonumber \\
  \hat T^a=d\hat e^a +\tfrac 12[\hat \omega^{ab}, \hat e_b]_\star -\tfrac i2 \epsilon_{abcd}\{\hat\omega_{bc},\hat f_c \}_\star+i\{\hat A^{(1)} , \hat e^a\}_\star -[\hat A^{(2)},\hat f^a]_\star, \nonumber\\
  \hat U^a=d\hat f^a +\tfrac 12[\hat \omega^{ab}, \hat f_b]_\star -\tfrac i2 \epsilon_{abcd}\{\hat\omega_{bc},\hat e_c \}_\star+i\{\hat A^{(1)} , \hat f^a\}_\star -[\hat A^{(2)},\hat e^a]_\star.\label{crvtrncmpnts}
\end{gather}
Up to f\/irst order,  the Seiberg--Witten maps for the component one forms $\hat \omega^{ab}$, $A^{(S)}$, $\hat e^a$ and $\hat f^a$ are
 \begin{gather}  \hat \omega^{ab}_\mu = \omega^{ab}_\mu +\tfrac 12\Theta^{\rho\sigma}\bigl(\omega^{ab}_\rho\bigl(\partial_\sigma A^{(1)}_\mu +F^{(1)}_{\sigma\mu}\bigr) + A^{(1)}_\rho \bigl(\partial_\sigma \omega^{ab}_\mu +R^{ab}_{\sigma\mu}\bigr)\nonumber\\
\phantom{\hat \omega^{ab}_\mu =}{}
+\tfrac 12\epsilon^{ab}_{\;\;\;\;cd} \omega^{cd}_\rho\bigl(\partial_\sigma A^{(2)}_\mu +F^{(2)}_{\sigma\mu}\bigr)+\tfrac 12\epsilon^{ab}_{\;\;\;\;cd} A^{(2)}_\rho\bigl(\partial_\sigma \omega^{cd}_\mu +R^{cd}_{\sigma\mu}\bigr) \bigr),\nonumber\\
  \hat A^{(1)}_\mu = A^{(1)}_\mu +\tfrac12 \Theta^{\rho\sigma}\bigl( A^{(1)}_\rho \bigl(\partial_\sigma A^{(1)}_\mu +F^{(1)}_{\sigma\mu}\bigr)- A^{(2)}_\rho \bigl(\partial_\sigma A^{(2)}_\mu +F^{(2)}_{\sigma\mu}\bigr)+\tfrac18[ \omega_{ab}]_{\rho} \bigl(\partial_\sigma \omega^{ab}_\mu +R^{ab}_{\sigma\mu}\bigr)\bigl),\nonumber\\
 \hat A^{(2)}_\mu = A^{(2)}_\mu +\tfrac12 \Theta^{\rho\sigma}\bigl(A^{(1)}_\rho \bigl(\partial_\sigma A^{(2)}_\mu +F^{(2)}_{\sigma\mu}\bigr)+A^{(2)}_\rho \bigl(\partial_\sigma A^{(1)}_\mu +F^{(1)}_{\sigma\mu}\bigr) \nonumber\\
 \phantom{\hat A^{(2)}_\mu =}{}
 -\tfrac 1{16} \epsilon_{abcd}\omega^{ab}_\rho\bigl(\partial_\sigma \omega^{cd}_\mu +R^{cd}_{\sigma\mu}\bigr) \bigr),\nonumber
\\
 \hat e^a_\mu = e^a_\mu +\Theta^{\rho\sigma} A^{(1)}_\rho \bigl(\partial_\sigma e^a_\mu +\tfrac 12 \omega^{ag}_\sigma e_{g\mu} - A^{(2)}_\sigma f^a_\mu\bigr) -\tfrac 14\Theta^{\rho\sigma}\epsilon^a_{\;\;bcd}\omega^{bc}_\rho  \bigl(\partial_\sigma f^d_\mu +\tfrac 12 \omega^{dg}_\sigma f_{g\mu} - A^{(2)}_\sigma e^d_\mu\bigr),\nonumber\\
 \hat f^a_\mu = f^a_\mu +\Theta^{\rho\sigma} A^{(1)}_\rho \bigl(\partial_\sigma f^a_\mu +\tfrac 12 \omega^{ag}_\sigma f_{g\mu} - A^{(2)}_\sigma e^a_\mu\bigr) \nonumber\\
 \phantom{\hat f^a_\mu =}{}
 -\tfrac 14 \Theta^{\rho\sigma}\epsilon^a_{\;\;bcd}\omega^{bc}_\rho  \bigl(\partial_\sigma e^d_\mu +\tfrac 12 \omega^{dg}_\sigma e_{g\mu} - A^{(2)}_\sigma f^d_\mu\bigr).
\label{swutfo}
\end{gather}

\subsection*{Acknowledgements}
I am very grateful to P.~Aschieri, S.~Fabi, B.~Harms and N.~Okada  for valuable discussions.

\pdfbookmark[1]{References}{ref}
\LastPageEnding

\end{document}